# A new class of fossil fragments from the hierarchical assembly of the Galactic bulge

F.R. Ferraro[1,2*], C. Pallanca[1,2], B. Lanzoni[1,2], C. Crociati[1,2], E. Dalessandro[2], L. Origlia[2], R.M. Rich[3], S. Saracino[4], A. Mucciarelli[1,2], E. Valenti[5,6], D. Geisler[7,8,9], F. Mauro[10], S. Villanova[7], C. Moni Bidin[10], G. Beccari[5]

**The formation and evolutionary processes of galaxy bulges are still unclear, and the presence of young stars in the bulge of the Milky Way is largely debated. We recently demonstrated that Terzan 5, in the Galactic bulge, is a complex stellar system hosting stars with very different ages and a striking chemical similarity to the field population. This indicates that its progenitor was likely one of the giant structures that are thought to generate bulges through coalescence. Here we show that another globular cluster-like system in the bulge (Liller 1) hosts two distinct stellar populations with remarkably different ages: only 1-3 Gyr for the youngest, 12 Gyr for the oldest, which is impressively similar to the old component of Terzan 5. This discovery classifies Liller 1 and Terzan 5 as sites of recent star formation in the Galactic bulge and provides clear observational proof that the hierarchical assembly of primordial massive structures contributed to the formation of the Milky Way spheroid.**

The picture of galaxy bulge formation is still largely debated in the literature[1]. Among the most credited scenarios, the "merging picture" proposes that galaxy bulges form from the merging of primordial substructures, either galaxies embedded in a dark matter halo[2,3], or massive clumps generated by early disk fragmentation[4-6]. Although the vast majority of the primordial fragments should dissolve to form the bulge, it is possible that a few of them survived the total disruption, and are still present in the inner regions of the host galaxy, grossly appearing like massive globular clusters (GCs). At odds with genuine GCs, however, these fossil relics should have been massive enough to retain the iron-enriched ejecta of supernova (SN) explosions, and possibly experienced multiple bursts of star formation. As a consequence, they are expected to host multi-iron and multi-age sub-populations. The first candidate fossil relic in the Galactic bulge was identified[7] back in 2009: the detailed photometric and spectroscopic study of Terzan 5 demonstrates[8-12] that this massive ($\sim 2 \times 10^6 M_\odot$) stellar system hosts at least two major sub-populations, ascribable to different star formation events. The first intense and short-lived (< 1 Gyr) star formation burst occurred about 12 Gyr ago, from gas enriched by type II SNe, with sub-solar metallicity ([Fe/H]= −0.3) and enhanced [α/Fe]=+0.4 abundance ratio. The second burst occurred ~7.5 Gyr later (approximately 4.5 Gyr ago), from gas characterized by super-solar metallicity ([Fe/H]=+0.3) and solar-scaled [α/Fe]. This suggests that the progenitor system of Terzan 5 was massive enough to retain a large amount of gas ejected from both type II and type Ia SNe, before igniting the second burst of star formation. The chemical abundance pattern of Terzan 5 is strikingly similar to that of the Galactic bulge[12], thus suggesting an intimate link between these two structures: Terzan 5 could be the remnant of one of the massive clumps that contributed to generate the Milky Way bulge in the framework of the proposed "merging picture". Of course, a sound confirmation of this possibility would come from the discovery of other similar fossil survivors of the Galactic bulge formation process.

We thus focused on another stellar system in the bulge, appearing as a GC, but still largely unexplored: Liller 1. It is located at only 0.8 kpc from the Galactic centre, very close to the Galactic plane (l= 354.84°, b= −0.16°)[13], in a region that is strongly affected by large foreground extinction[14,15]. Indeed, with an average colour excess E(B−V) larger than 3, an extinction as large as 10 magnitudes can be estimated in the optical band toward the cluster. This, combined with significant evidence of differential reddening[14], has severely hampered the study of Liller 1 so far. A first insight into its stellar content was recently obtained[16] from a dedicated study based on near-IR GEMINI observations, which revealed a very large total mass (>2 $10^6$ $M_\odot$), similar to that of Terzan 5, and provided the accurate determination of the structural and physical parameters of the system (scale radii, concentration parameter, central mass density).

Here we present the very first ultra-deep observations of Liller 1 in the V and I bands, by using the ACS-WFC on board the HST (see Section 'Data-set and Analysis' in Methods and Extended Data Figure 1). These observations have been combined with the near-IR (J and K) images obtained with GEMINI to select the most suitable colour-magnitude diagram (CMD) for the study of the stellar populations in Liller 1. The analysis shows that the best images (in terms of both spatial resolution and limiting magnitudes) are those acquired through the I and the K filters.

Figure 1a shows the (I, I−K) CMD obtained from these observations, sampling a 90"x90" region roughly centred on the centre of gravity[16] of the stellar system. In spite of the evident effect of differential reddening, the main evolutionary sequences are defined and the analysis of the CMD reveals a few major intriguing features: (1) a sparse stellar population defining a sequence at the blue edge of the CMD, (2) a dominant population typical of an old star cluster, consisting of cool and bright stars at the red side of the CMD, at (I−K) ~6.5, joining a rich main sequence (MS) extending down to I~26.5, and (3) a prominent Blue Plume (hereafter BP) at (I−K)~5.7, appearing as an extension of the old MS, which was also visible in the published[16] IR CMD at (J-Ks)~1.8 and 15.5<Ks<17.5. While the first component is related to the MS of the Galactic disk (see below; we thus name it Disk-Field population), the dominant one shows all the properties typical of an old metal-rich component, with a well populated Red Giant Branch (RGB) extending for more than 6 magnitudes and a well-defined Red Clump (RC) at I~21. Hereafter, we refer to this component as Old Population (OP). The most astonishing feature in the CMD is the BP, which is very populated, extends for approximately 4 magnitudes, and flows with no discontinuity into the MS of the OP: indeed, such a component is totally unexpected in the CMD of an old GC.

To understand the nature of this component, we determined the star density profile (see Section 'The sub-population star density profiles' in Methods and Extended Data Fig.2) of the three sub-populations selected in Figure 1a, with respect to the Liller1 centre[16]: as shown in Figure 1b, while the Disk-Field population shows the standard flat profile expected for a pure field population, the BP and the OP appear highly concentrated toward the cluster centre, designing a well peaked profile nicely reproduced by the same King model (solid lines). This clearly indicates that, although affected by some level of field contamination (manifested by the flattening of the profile at large distances r>100"), both populations belong to Liller1.

To properly separate cluster members from Galactic field interlopers, we determined the stellar proper motions (PMs) from the available observations, which sample a temporal baseline of 6.3 years (see Section 'Proper motions' in Methods). As shown in Figure 2a,b, the Disk-Field population is completely removed by the PM analysis and the level of contamination from field stars along the OP and the BP is appreciable from the comparison between the field- and the cluster member-CMDs. The cluster member CMD shows the indisputable presence of the BP, fully confirming that this population does belong to Liller 1.

Figure 2c shows the vector-point diagram of the three sub-populations selected from the boxes depicted in Fig. 1a. Interestingly, the Disk-Field and the OP distributions are very different, and the BP appears to host both these components: a main population with kinematics perfectly consistent with that of the OP, and another component with elongated PM distribution similar that of the Disk-Field. The two components of the BP are well distinguished by the PM analysis, confirming beyond any doubt that the majority of the BP stars do belong to Liller1.

The analysis presented in the rest of the paper is referred to the sample of stars (~80% of the total) for which PMs have been measured.

To more precisely define the evolutionary sequences in the CMD, we corrected for the effects of differential reddening, which is highly variable on small scales in the direction of Liller 1. We constructed a high spatial-resolution extinction map following the approach already adopted in previous works[17] (see Section 'The extinction map in the direction of Liller 1' in Methods). The CMD of the PM-cleaned and differential reddening corrected (DRC) stellar population of Liller1 is shown in Figure 3a. It clearly confirms the complexity of the system: Liller1 hosts an ensemble of stars not following the evolutionary sequences typical of an old stellar population. This BP appears very well populated, counting a number of objects comparable to that of the OP in the same magnitude range (see Sections 'BSSs or not BSSs' and 'Artificial star experiments and completeness' in Methods; see also Extended Data Fig. 3). The cumulative radial distributions of the PM-cleaned samples (Figure 3b) fully confirm the conclusions above: following the Kolmogov-Smirnov test, the probability that the BP and the OP are extracted from the parent distribution describing the Disk-Field population is zero. Indeed, the BP appears even more centrally segregated toward the gravitational centre of Liller1 than the OP, once more demonstrating its secure membership to the system and providing some additional insights on its origin (see below).

**What is the origin of the BP population? -** The BP population lies in the region of the CMD usually occupied by Blue Straggler Stars (BSSs)[18]. However, while BSSs commonly appear as a sparse population in the CMD, the number of stars counted along the BP is of the same order of that along the OP in the same magnitude range (see Section 'BSSs or not BSSs' in Methods). Indeed, the richness and the morphology of the BP population are more suggestive of a coherent and recent episode of star formation, than a population of BSSs.

The next step therefore is the dating of the two sub-populations of Liller 1, which also requires the knowledge of their metallicity. Here we adopted a provisional global metallicity of [M/H]=−0.3 (corresponding to [Fe/H]=−0.3 with no α-enhancement, or, e.g., [Fe/H]=−0.4 and [α/Fe]~+0.2) for the bulk population of Liller1 (see Section 'Liller1 metallicity ' in Methods). By adopting PARSEC isochrones[19,20] with ages between 10 to 14 Gyr, we found that the RGB colour and the magnitude level of the Red Clump of the OP are best reproduced with E(B-V)= 4.52±0.10 (which is larger than previous estimate, but see Section 'The extinction map in the direction of Liller 1' in Methods) and a distance modulus $(m-M)_0$=14.65±0.15 (consistent with the latest determination[16]). With these values, the morphology of the MS-turnoff and sub-giant branch (SGB) provides a best-fit age of t=12 ±1.5 Gyr for the OP (see Figure 4, Section 'Measuring the age of the old stellar population' in Methods and Extended Data Fig. 4). This finding confirms that Liller 1 harbours a very old stellar component, formed during the remote epoch of the Milky Way formation.

The accurate dating of the BP population is a much more difficult task. In fact, from the analysis of the CMD, it is evident that significantly younger isochrones are needed to reproduce the BP, but no metallicity measurements for stars along this sequence are available at the moment. While the simpler assumption is to adopt the OP metallicity also for the BP, a self-enrichment scenario would predict an increased metal content for the younger population. We thus explored the following two reference scenarios: (1) the stars in the two sub-populations (BP and OP) share the same metallicity ([M/H]=−0.3), and (2) the BP is more metal-rich than the OP, with iron and α-element abundances consistent with those observed in Terzan5 ([Fe/H]=+0.3 and [α/Fe]=0)[8,11]. Figure 5a summarizes the results. For illustrative purposes, three isochrones (of 1, 2 and 3 Gyr) are superimposed on the CMD, by assuming the OP metallicity (left panel), and a super-solar Terzan5-like iron content (right panel). As can be seen, the brightest portion of the BP turns out to be much better reproduced by the super-solar, 1 Gyr-old isochrone. In any case, independently of the exact value of the metal content, both scenarios show that the distribution of stars observed in the BP requires the presence of a population as young as 1-Gyr!

While the presence of a centrally segregated young population (see Figure 3b) is hard to explain in a scenario where Liller 1 comes from the merger of two star clusters (which is strongly disfavoured also by the fact that young and massive stellar systems are very rare in the Milky Way), it is well consistent with an auto-enrichment process, where a new generation of (more metal-rich) stars formed in the central regions of the system and possibly further sank toward the centre because of mass segregation (in fact, the central relaxation time[13] of Liller 1 is of the order of $10^6$ yr only and, according to their age, these stars should be a factor of 2-2.5 more massive than the MS-turnoff mass of the OP). Indeed, the preliminary comparison with synthetic CMDs indicates that the observed BP is accurately reproduced in a self-enrichment scenario, with a second star formation burst starting ~3 Gyr ago and lasting until 1 Gyr ago (see Figure 5b and Section 'Synthetic CMD' in Methods).

These results clearly classify Liller 1 as a non-genuine GC and collocate it in the class of GC-like structures in the Galactic bulge harbouring multi-age stellar components, which was initiated by the discovery of the complex stellar populations in Terzan 5.

**Liller1 and Terzan5: differences and similarities -** The comparison of the stellar populations in these two stellar systems is quite instructive. In Figure 6, their CMDs have been aligned by requiring that the faintest RC of Terzan 5 matches that of Liller 1. The comparison shows both gross differences and some relevant similarities in the properties of the stellar populations in these two systems.

*Differences* − The main difference between the two CMDs concerns the morphology of the young population. In Terzan 5 the major manifestation of this component is the presence of a RC brighter and redder than that of the old population. In Liller 1 no relevant features are detected in the evolved portion of the CMD and the young population manifests its own presence in the form of an extended MS, suggesting a quite recent burst of star formation (occurred ~1-3 Gyr ago). The post-MS evolutionary timescale of such young stellar population is so fast that no relevant features are expected to be observable in the evolved portion of the CMD, apart from the RC. Theoretical isochrones[19,20] indicate that the RCs of the young and old components overlap in the CMD (see Figure 5a), and should therefore be indistinguishable in the diagram. However, the existence of the BP should produce a numerical excess with respect to the amount of RC stars expected if only the OP was present. Star counts confirm that this is indeed the case (see Section 'Star excess in the RC region' in Methods).

*Similarities* − The comparison of the two CMDs also shows a striking similarity between the two old stellar populations. Indeed, if the RCs of the two systems are aligned, also their RGB–bumps, SGBs and the MS-TO points appear to be at the same magnitude levels. In other words, the old populations in the two stellar systems are strikingly similar and they are reproduced by the same isochrone (with t=12 Gyr and [M/H]= −0.3). This finding suggests that both stellar systems formed at the epoch of the Galaxy formation (~1 Gyr after the Big-Bang), from gas clouds with similar chemistry. The detailed study of the α-element abundance patterns of the OP in Liller 1 is needed to confirm this hypothesis.

**A new class of objects -** The properties of the old stellar populations observed in Liller1 and Terzan 5 demonstrate a concurrent genesis of these two systems, collocating them at the early epoch of the Galaxy assembly. The discovery[21,22] of an increasing number of stellar streams, interpreted as the relics of cannibalism events experienced by the Milky Way in its past history, may suggest that Terzan 5 and Liller 1 are nuclear star clusters of massive satellites accreted *after* the epoch of the Galaxy formation. However, the abundance patterns observed for the old stellar population in Terzan 5 clearly indicate that it formed from a high metallicity (half-solar) cloud enriched exclusively by the ejecta of type II SNe. This chemical signature is strikingly similar to that observed for the bulge field stars and, in fact, it is typical of massive and dense environments that experienced star formation at very high rates, as the bulge. It therefore seems very unlikely that such an event could occur in a low-mass satellite galaxy, and strongly supports, instead, an *in situ* origin of Terzan 5. Although the complete chemical characterization (in terms of α-elements) of the OP of Liller 1 is still lacking, its similarity with Terzan 5, in terms of location (the bulge), age, and (likely) iron content, strongly supports a common origin for these two systems, and their common link with the Galactic bulge. In fact, a scenario where, just by chance, the Milky Way bulge and the nuclear star clusters of two different satellite galaxies independently followed the *same* chemical enrichment process and are now all observed in the *same* region of the Milky Way is, in our view, quite unlikely. Conversely, the age and similarity observed for the old populations in these two systems suggest that both Terzan 5 and Liller 1 are primordial structures that formed in situ from highly metal-enriched gas, as expected in the environment where the proto-bulge assembled.

The differences between the young populations detected in the two systems indicate that the subsequent star formation events were activated and occurred over quite different time-scales, possibly tracing different histories of strong interactions with the local environment or other sub-structures, also depending on the respective orbits within the Galactic potential. On the other hand, the properties of the young populations, which appear to be more metal-rich and more centrally segregated than the old populations (in agreement with what is expected in a self-enrichment scenario), confirm that these systems were massive enough to retain the SN ejecta. Thus, the combination of the available observational evidence defines these structures as the likely remnants of primordial massive systems that formed in situ and contributed to generate the bulge ~12 Gyr ago: this is a new class of objects, that here we name Bulge Fossil Fragments (BFFs). The detection[7,11] of similar structures in the star forming regions of high-redshift galaxies confirms that such massive fragments existed at the epoch of the Milky Way assembly. By using the mass-metallicity relation[23] for galaxies at z=3-4, an initial mass of a few $10^9$ $M_\odot$ is expected[12] for these BFFs. Thus, about ten of such primordial massive structures could have provided the entire mass budget (~2 $10^{10}$ $M_\odot$)[24] of the Galactic bulge[25]. Most of these structures merged together to form the bulge, but, for unknown reasons, a few of them survived the total disruption[16] and are now observable as BFFs. In this respect their discovery provides the first observational proof that hierarchical assembly played a role also in the formation of the Galactic bulge. Thus, these findings, on one side reconcile[12] the Milky Way bulge formation process with the observations at high-z, on the other side, they suggest that bulge formation has been more complex than previously though, with a significant contribution from both the merger of pre-formed, internally-evolved stellar systems and the formation and secular evolution of the bar and other sub-structures[26].

**Conclusion and future perspectives -** The discovery reported in this paper allowed us to define a new class of stellar systems orbiting the Galactic bulge. These systems (1) are indistinguishable from genuine GCs in their appearance, (2) have metallicity and abundance patterns compatible with those observed in the bulge field stars, (3) host an old stellar population (testifying that they formed at an early epoch of the Galaxy assembling), (4) host a young stellar population, several to many Gyrs younger than the old one (testifying their capacity of triggering multiple events of star formation). The detailed characterization of the chemical patterns of the OP and the BP population in Liller 1 is now urged to firmly constrain the formation history of this BFF.

The survival of a few BFFs up to now suggests that they evolved in an environment significantly less violent than the forming bulge. Their properties also demonstrate that for several Gyrs they retained a large amount of gas ejected by SN explosions, before producing a new generation of stars; such bursty star formation histories, with long periods of quiescence, are not rare in the Universe[27] and in the Galaxy. In particular, multiple star-formation episodes have been recently found to characterize the star formation history of the Galactic Disk[28] and Galactic Center[29], possibly related to gravitational interactions and the accretion of small satellite galaxies (as the Sagittarius dwarf galaxy). The epoch of the most recent star formation burst appears to differ from BFF to BFF, for reasons that need to be further investigated, but are possibly related to major interactions with other bulge sub-structures or the Galactic bar. In this respect, reconstructing the orbit of these systems[30] can bring new crucial information. Such interactions could also be responsible for the severe mass-loss suffered by these structures during their evolution, that significantly reduced their original mass ($\sim 10^9\,M_\odot$) to the current value ($\sim 10^6\,M_\odot$)[9,16].

The discovery that Liller 1 belongs to this class of objects confirms the prediction[7,12] that other surviving remnants could still be hidden in the heavily obscured regions of the Milky Way bulge. Thus a systematic search for other candidates is crucial in the next years. Finally, our results definitely identify the BFFs as sites of recent star formation in the Galactic bulge, suggesting that the still debated fraction of young stars in the bulge field[31,32] could come from recent destruction or stripping processes involving some of the survived BFFs.

Correspondence and requests for materials should be addressed to Francesco R. Ferraro (email: francesco.ferraro3@unibo.it).

**Acknowledgements:** This research is part of the project COSMIC-LAB at the Physics and Astronomy Department of the Bologna University (see the web page: http://www.cosmic-lab.eu/Cosmic-Lab/Home.html) at the University of Bologna. The research has been funded by project *Light-on-Dark*, granted by the Italian MIUR through contract PRIN-2017K7REXT. The research is based on data acquired with the NASA/ESA HST under project GO-15231 (PI Ferraro) at the Space Telescope Science Institute, which is operated by AURA, Inc., under NASA contract NAS5-26555. Based also on data secured at the GEMINI South Telescope. SS gratefully acknowledge financial support from the European Research Council (ERC-CoG-646928, Multi-Pop). EV acknowledges the Excellence Cluster ORIGINS Funded by the Deutsche Forschungsgemeinschaft (DFG, German Research Foundation) under Germany´s Excellence Strategy – EXC-2094 – 390783311. D.G. gratefully acknowledges support from the Chilean Centro de Excelencia en Astrofísica y Tecnologías Afines (CATA) BASAL grant AFB-170002. D.G. also acknowledges financial support from the Dirección de Investigación y Desarrollo de la Universidad de La Serena through the Programa de Incentivo a la Investigación de Académicos (PIA-DIDULS). SV gratefully acknowledges the support provided by Fondecyt reg. n. 1170518.

**Author information**

**Affiliations**

Dipartimento di Fisica e Astronomia, Università di Bologna, Via Gobetti 93/2, I-40129 Bologna, Italy

Francesco R. Ferraro, Cristina Pallanca, Barbara Lanzoni, Chiara Crociati & Alessio Mucciarelli

INAF -- Astrophysics and Space Science Observatory Bologna, Via Gobetti 93/3, I-40129 Bologna, Italy

Francesco R. Ferraro, Cristina Pallanca, Barbara Lanzoni, Chiara Crociati, Emanuele Dalessandro, Livia Origlia & Alessio Mucciarelli

Department of Physics and Astronomy, 430 Portola Plaza, Box 951547, Los Angeles, CA 90095-1547, USA

R. Michael Rich

Astrophysics Research Institute, Liverpool John Moores University, 146 Brownlow Hill, Liverpool L3 5RF, UK

Sara Saracino

European Southern Observatory, Karl-Schwarzschild-Straβe 2, D-85748 Garching bei München, Germany

Elena Valenti & Giacomo Beccari

Excellence Cluster ORIGINS, Boltzmann-Straβe 2, D-85748 Garching bei München, Germany

Elena Valenti

Departamento de Astronomía, Universidad de Concepción, Casilla 160 C, Concepción, Chile

Douglas Geisler & Sandro Villanova

Instituto de Investigación Multidisciplinario en Ciencia y Tecnología, Universidad de La Serena. Avenida Raúl Bitrán S/N, La Serena, Chile

Douglas Geisler

Departamento de Física y Astronomía, Facultad de Ciencias, Universidad de La Serena. Av. Juan Cisternas 1200, La Serena, Chile

Douglas Geisler

Instituto de Astronomía, Universidad Católica del Norte, Av. Angamos 0610, Antofagasta, Chile

Francesco Mauro & Cristian Moni Bidin

**Contributions**

FRF designed the study and coordinated the activity. CP, CC, BL, ED, AM, SS, EV, FM, SV, CMB and GB analysed the photometric datasets. ED determined the stellar proper motions. FRF, CP and BL wrote the first draft of the paper. ED, LO, RMR, EV and DG critically contributed to the paper presentation. All the authors contributed to the discussion of the results and commented on the manuscript.

**Corresponding author**

Correspondence to Francesco R. Ferraro

**Ethic declarations**

Competing Interests

The authors declare no competing interests.

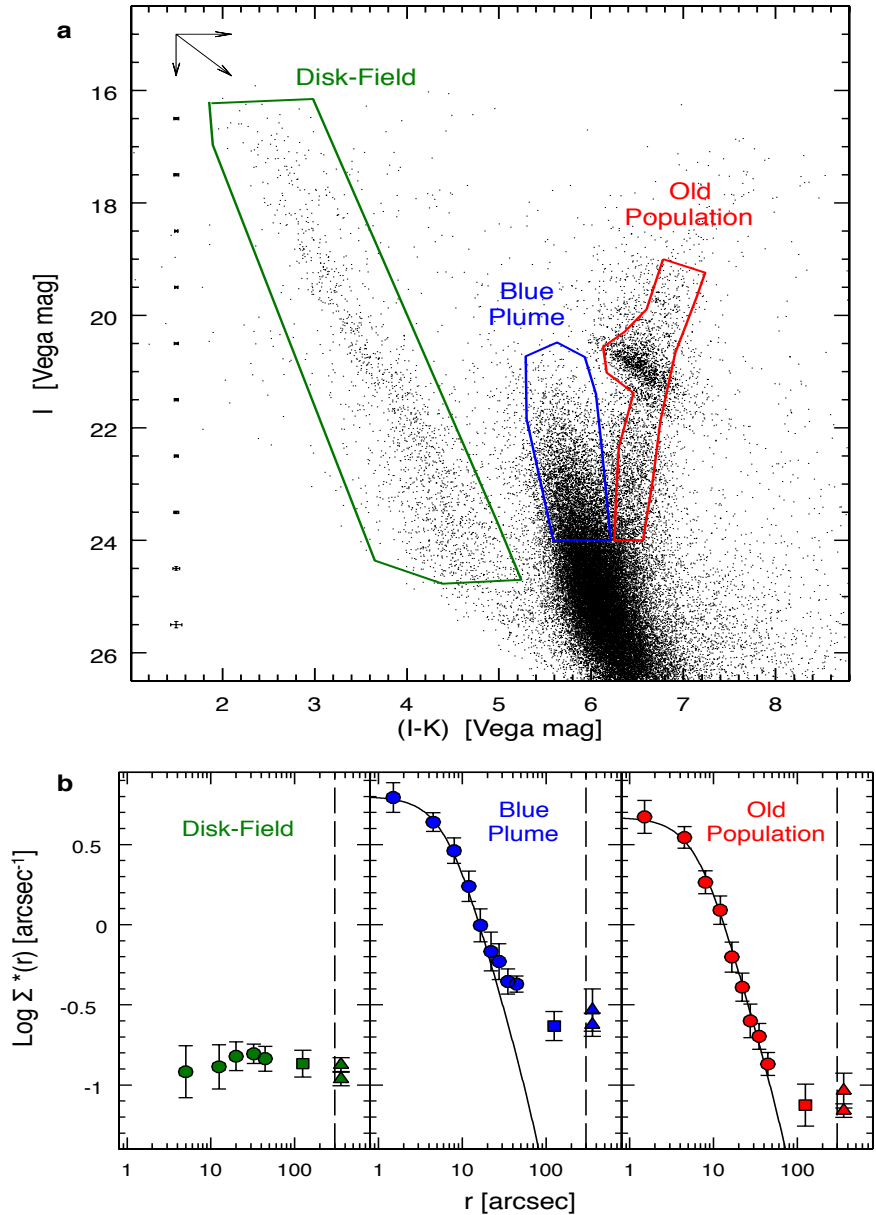

**Figure 1 – The stellar sub-populations identified in the CMD of Liller 1. a,** Optical/near-IR colour-magnitude diagram of Liller 1, approximately sampling the 90"x90" central region of the system, as obtained from the combination of HST-ACS and GeMS-GEMINI observations. Three main sub-populations can be distinguished: (1) a sparse blue population at (I-K)<5 that is related to the main-sequence of the Galactic Disk (thus it is named Disk-Field population), indicated by the green box; (2) a group of bright red stars, with (I-K)~6.5, that joins a well populated MS extending from I~24.5 down to I~26.5, tracing an old metal-rich component ("Old Population, OP"), marked with the red box; and (3) a peculiar Blue Plume (BP) tracing a sub-population of young stars, evidenced with a blue box. The mean errors (1 s.e.m.) in magnitude and colour are reported on the left for 1 magnitude-wide bins. The reddening vector is also shown. **b,** Star density profiles of the three sub-populations identified in the direction of Liller 1. The coloured circles, squares and triangles correspond to the stellar density values measured in the HST-GEMINI (r<50"), NICMOS (110"<r<140"), and WFC3 (r>330") fields of view, respectively. The error bars refer to the 1σ uncertainty. The vertical shaded line marks the tidal radius of Liller 1. The solid line is a King model with core radius rc=7".

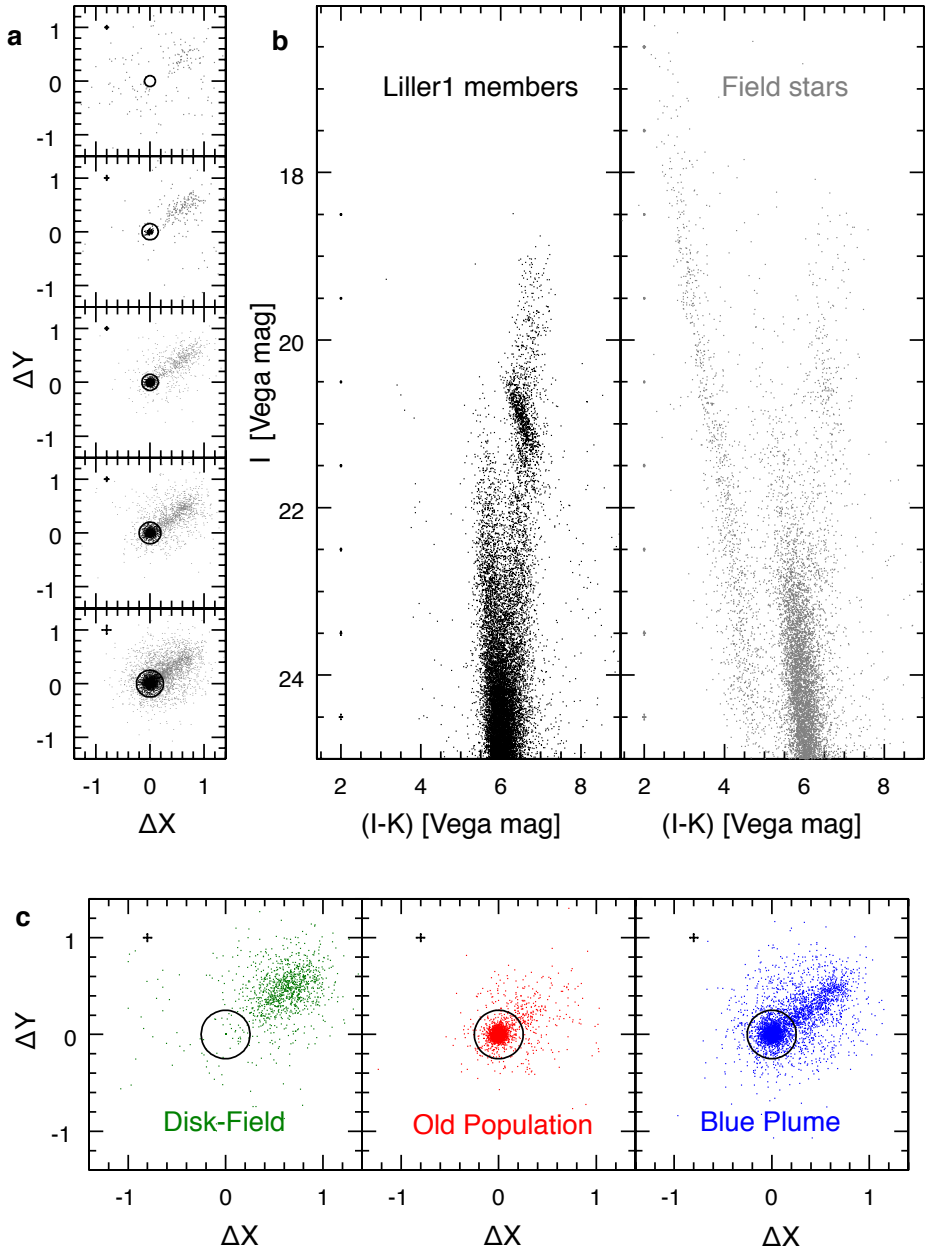

**Figure 2 – Separating Liller 1 members from Galactic field interlopers: proper motion (PM) analysis. a,** Vector-point diagrams (VPD) of the relative PMs (in units of the ACS/WFC pixel, corresponding to 0.05") measured for stars in five magnitude bins. The stars included within the circles centred on (0,0) are selected as cluster members (black dots), while those outside the circles are considered as field interlopers (grey dots). The radius of each circle corresponds to 3 times the typical PM error in the magnitude bin (which is marked with the crosses shown in each panel). **b,** CMDs of Liller 1 members (left, in black) and field stars (right, in grey) as obtained from the PM analysis. Both panels show the mean errors (1 s.e.m.). **c,** VPDs for the three sub-populations (Disk-Field, OP, BP) selected by adopting the boxes shown in Fig.1a. The Disk-Field component shows an elongated PM distribution avoiding the central portion of the VPD. Conversely, a circularly symmetric distribution around (0,0) is observed (as expected) for the OP.

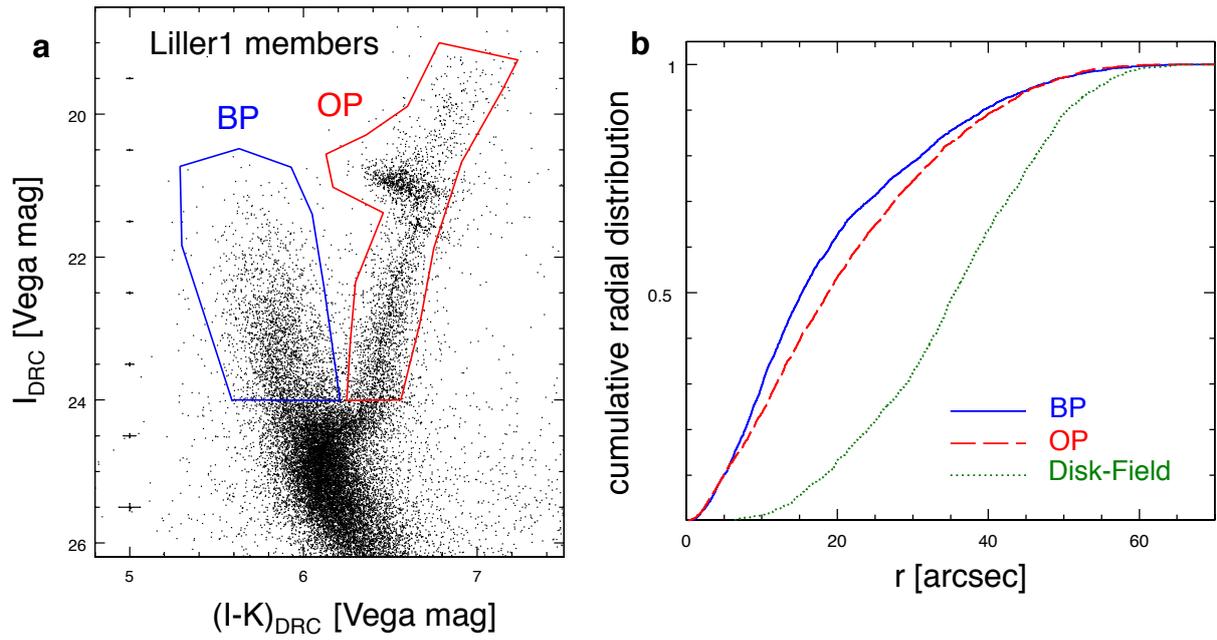

**Figure 3 – The properties of the Liller 1 stellar populations. a**, CMD of the PM-selected members of Liller 1 with magnitudes ($I_{DRC}$) and colours (($I-K)_{DRC}$) corrected for differential reddening. The selection boxes of the BP (in blue) and OP (in red) used for the study of their radial distribution are also shown. The mean errors (1 s.e.m.) in magnitude and colour are reported on the left for 1 magnitude-wide bins. **b**, Cumulative radial distributions of the three sub-populations observed in the PM-selected CMDs of Liller 1. The numbers of stars counted in the selected components are: 2480 in the BP, 2916 in the OP, and 1109 in the Disk-Field.

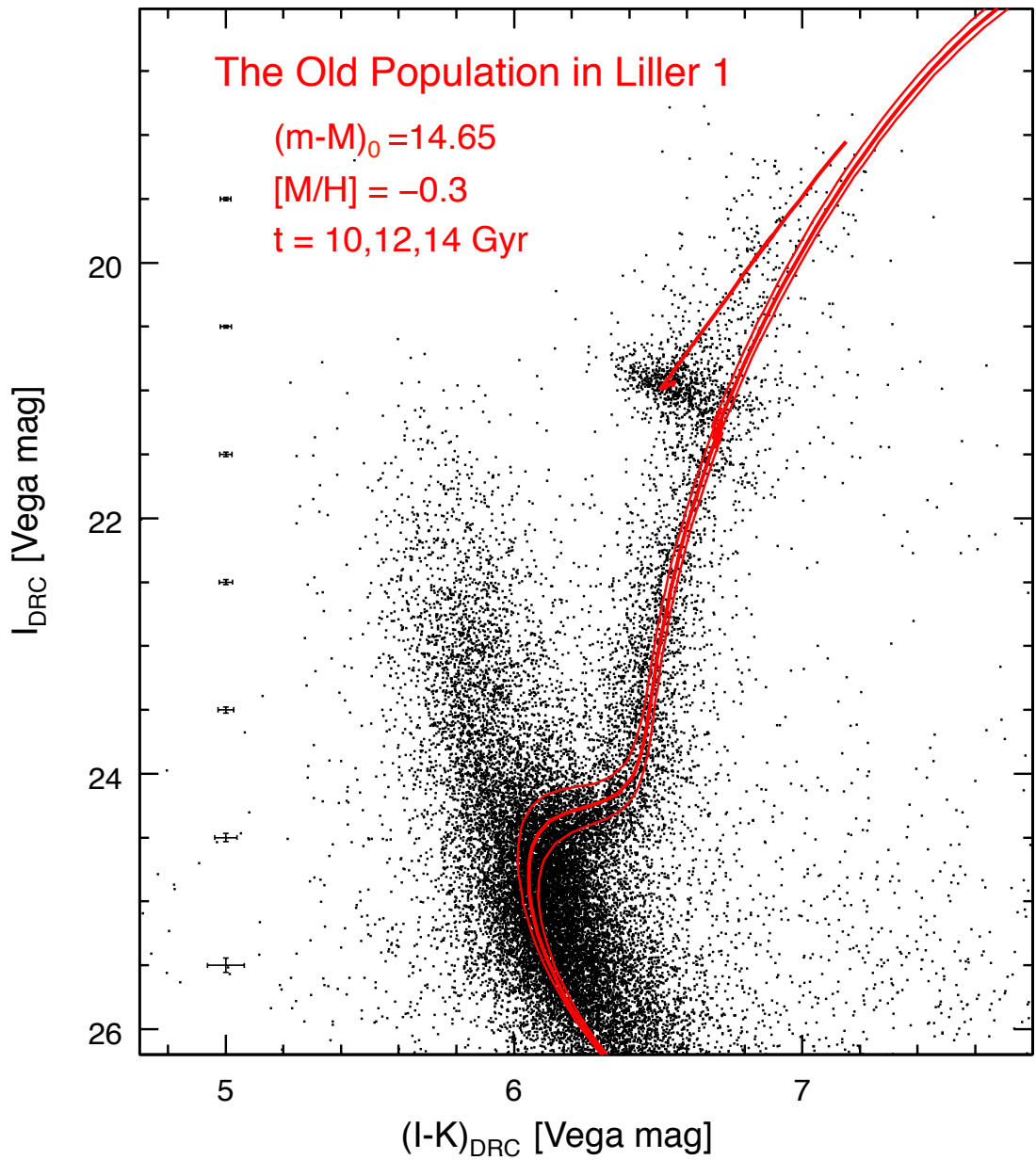

**Figure 4 – The age of the old stellar component in Liller1.** Differential reddening corrected CMD of Liller 1 with three isochrones[19,20] of different ages (10, 12 and 14 Gyr, from top to bottom) superimposed in red. The adopted distance modulus and metallicity are also labelled. The mean errors (1 s.e.m.) in magnitude and colour are reported on the left for 1 magnitude-wide bins.

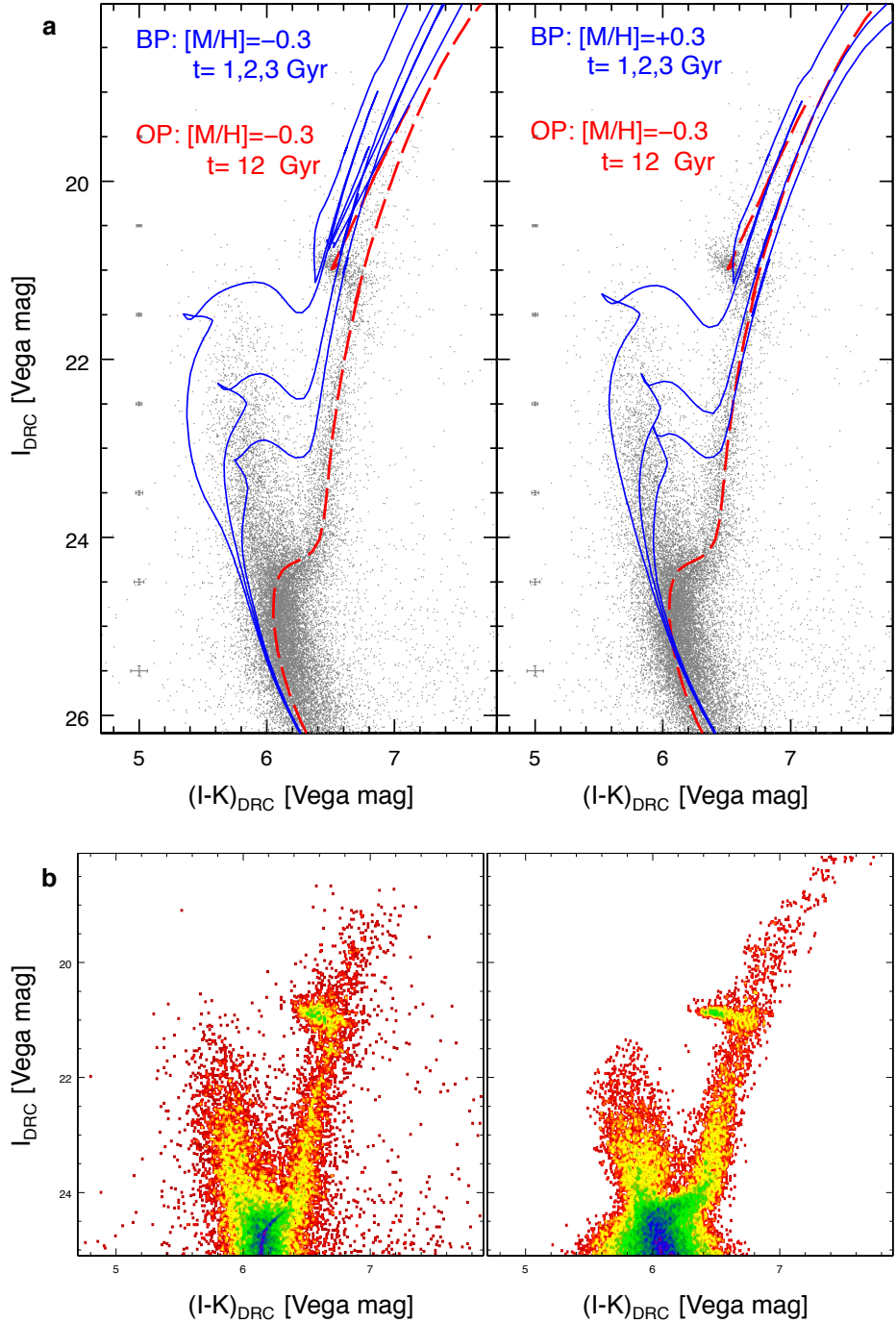

**Figure 5 – The complex stellar content of Liller1. a,** Three young isochrones[19,20] (of 1, 2 and 3 Gyr, solid blue lines from top to bottom) are over-plotted to the CMD of Liller 1, where also the best-fit isochrone of the OP is shown (red dashed line). Since the metal content of the BP is currently unknown, for the sake of illustration, the isochrones have been calculated by assuming the same metallicity of the OP (left panel), [M/H]=−0.3, and a significantly larger value, [M/H]=+0.3 (right panel). The mean errors (1 s.e.m.) are reported. **b**, Comparison between the CMD of the PM-selected members of Liller1 (left) and a synthetic CMD (right) obtained by combining a 12-Gyr old stellar population with a set of young populations with ages in the range 1-3 Gyr. The stellar density in the CMD is colour-coded as follows: from blue, to green, yellow and red, for decreasing density.

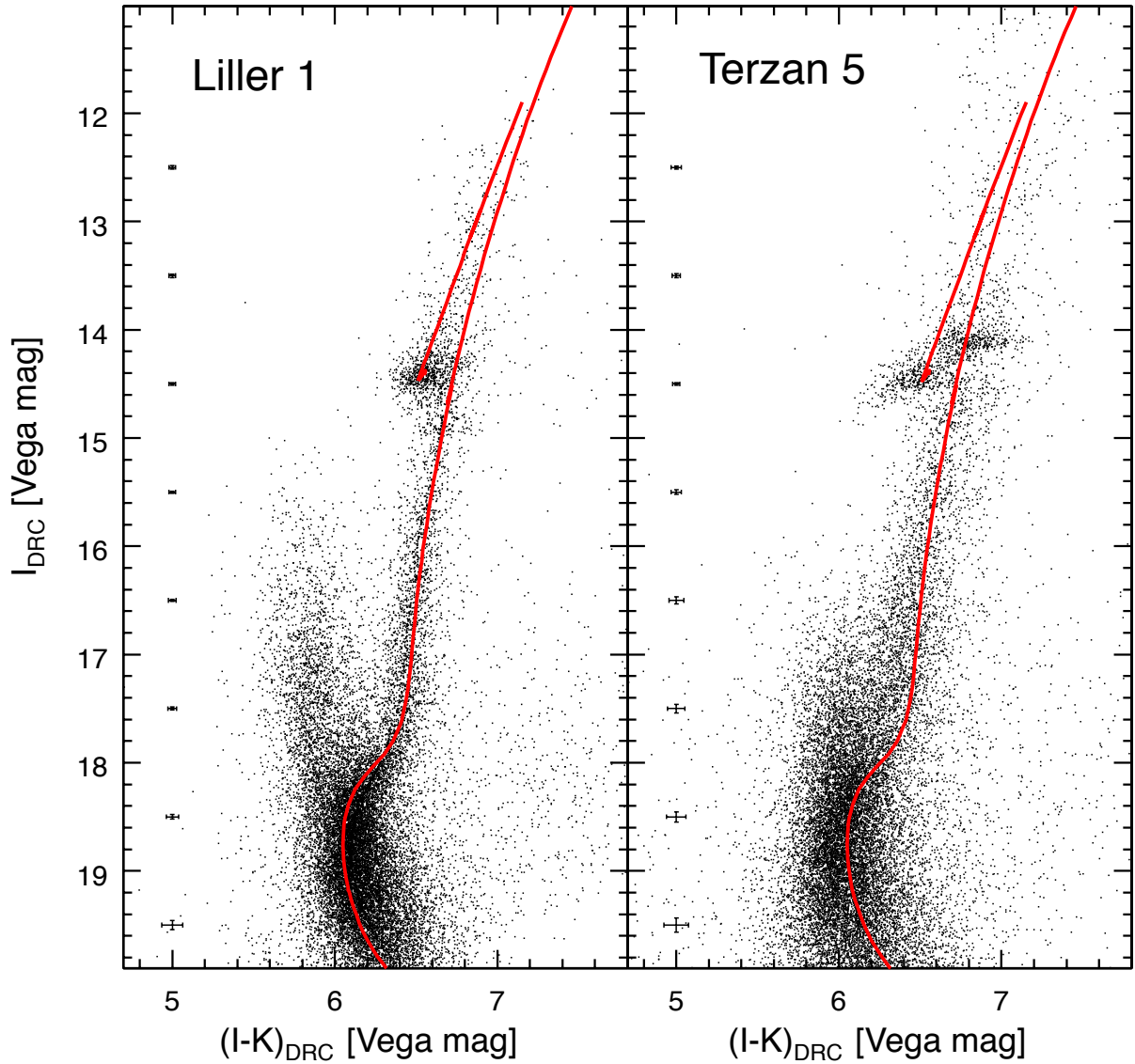

**Figure 6– Liller1 and Terzan 5: two primordial fragments originated from similar gas clouds. a,** CMD of Liller 1 with the isochrone[19,20] used to estimate the age of its OP (t=12 Gyr and [M/H]=−0.3) superposed as a red line. **b,** CMD of Terzan 5 shifted in magnitude and colour to make its old (faintest) Red Clump match the Red Clump of Liller 1, with the same isochrone shown in panel **a** superposed in red. The reported errors are 1 s.e.m.

## Methods

**Data-set and Analysis:** For this study we have secured the first set of space-based high-resolution optical images ever acquired for Liller 1, by using the Wide Field Channel of the Advanced Camera for Survey (ACS/WFC) on board the Hubble Space Telescope (GO15231, PI: Ferraro). The observations have been performed in the filters F606W (V) and F814W (I) and consist in a set of 6 deep images per filter. We also used near-IR observations secured with the adaptive-optics assisted camera GeMS/GSAOI at the GEMINI south Telescope. The details of the observations are provided in Extended Data Fig. 1.

The photometric analysis of the optical dataset was performed via the point-spread function (PSF) fitting method, by using DAOPHOT II[33] following the "standard" approach used in previous works[34]. Briefly, spatially variable PSF models were derived for each image by using some dozens of stars, and then applied to all the sources with flux peaks at least 3 σ above the local background. We then built a master-list containing all the sources detected in more than 3 images in at least one filter. At the position of each star belonging to the optical master-list, we then forced a fit in each frame of both filters, by using DAOPHOT/ALLFRAME[33,35]. For each star thus recovered, multiple magnitude estimates from different exposures were homogenised by using DAOMATCH and DAOMASTER, and their weighted mean and standard deviation were finally adopted as star magnitude and photometric error. The final HST catalogue contained all the stars measured in at least 3 images in any of the two filters. The instrumental optical magnitudes were calibrated onto the VEGAMAG photometric system[36] by using the updated recipes and zero-points available in the HST web-sites. The instrumental coordinates were first corrected for geometric distortion and then reported to the absolute coordinate system (α, δ) by using the stars in common with the publicly available Gaia DR2 catalog[37]. The resulting 1σ astrometric accuracy is typically ≤ 0.1".

The analysis of the optical catalogue demonstrated that, because of the large extinction in the direction of the system, the V-band exposures are significantly shallower that those secured through the I filter. A similar situation affected the IR catalog[16]: the J images are much shallower than those obtained through the K filter. This clear indicated that the optimal solution for the accurate study of the stellar populations in Liller 1 consisted in combining the deepest available images, namely those acquired in the I and in the K bands.

For this reason, we proceeded with a second-step analysis consisting in a combined analysis of the I and K exposures. Starting from the first step of the independent reduction of the two datasets, two lists of sources have been obtained: one comprising all the stars detected in the I filter, and one including the objects identified in the K band. The I-band list contained all the stars measured in at least 3 frames, over the 6 available. The K-band list contained all the stars measured in at least 3 over the 6 best-quality images analysed. IR photometry for the brightest stars (K<12.5), which are saturated in the GEMINI observations, has been taken from a catalog[16] derived from the VISTA Variables in the Vía Láctea Survey (VVV)[38].

We reported the K-band list onto the I-band one, and we built a single master-list containing all the detected stars (some having magnitude measured in both filters, some other having only the I or the K magnitude). In this way, sources detected only in the I images where then searched and analysed in the K images and viceversa, thus maximizing the information provided by the two filters separately. The final magnitudes of this combined analysis were homogenized to the calibrated stand-alone HST and GeMS catalogues.

The final catalogue counts 43658 stars, reaching I~26.5 with signal-to-noise ratio S/N~8.

Parallel observations with the HST-WFC3 camera, in the F606W (V), F814W (I), F110W (J) and F160W (H) filters have been also secured in order to sample the Galactic field population in the region surrounding the cluster. Two "control fields" have been observed. They are located at ~5' from the centre of Liller 1 (roughly corresponding to the tidal radius of the system)[16] in the SE and NW directions. These observations have been analysed following the procedure described above. The (I, I-H) CMDs of the external portion (r>330") of the two WFC3 control fields are shown in Extended Data Figure 2 (panels a,b). They clearly show that a different level of extinction is affecting the SE and the NW fields. Moreover, the comparison with Figure 1a confirms that, in spite of the colour difference, Galactic field stars are observed in the same CMD locations where the OP and the BP are selected (i.e., both the OP and the BP populations are contaminated by field components).

To also sample a radially intermediate region, we retrieved NICMOS observations (GO7318) from the HST Archive, and analysed a 50"x 50" field approximately located at 100" from the cluster centre. Because of the relatively low spatial resolution of NICMOS (the pixel scale is 0.2"), we used the stars detected in the ACS-WFC catalogue as input for the NICMOS analysis. From the combination of the I-band ACS-WFC magnitudes, and the H-band NICMOS magnitudes, we obtained the (I, I-H) CMD shown in Extended Data Figure 2 (panel c).

**The sub-population star density profiles:** We determined the star density profile of the three sub-populations detected in the direction of Liller1 by following the same procedure already adopted in previous papers[10,39-41]. In short, we divided the HST-GEMINI field in 9 concentric annuli centred on the gravitational centre of the system, each one split in (typically) four sub-sectors. For the three selected sub-populations (see Figure 1a), we then counted the number of stars lying within each sub-sector and divided it by the sub-sector area. The stellar density in each annulus was finally obtained as the average of the sub-sector densities, and the standard deviation among the sub-sectors densities was adopted as error. The resulting stellar density profiles $\Sigma_*(r)$, in units of number of stars per square arcsecond, are shown (circles) in Figure 1b. To investigate the density profile at larger distances from the centre and thus get an estimate of the field contamination level for each sub-population, we applied the same procedure to the WFC3 NE and SW control fields, and to the NICMOS data (see the section 'Data-set and Analysis'). In these cases, however, for adopting the same sub-population selection boxes, we had first to transform the observed (I-H) into the (I-K) colour. To this end, we used the stars in common with the VVV survey[38]. Because of the significantly different reddening affecting the two WFC3 pointings, we also shifted the CMD of the SE control field to match that of the NW field (which is more similar to that of Liller1). The stellar density measured in the NICMOS field for the three sub-populations is shown with square symbols in Figure 1b, while the results obtained in the WFC3 control fields for r>330" are marked with triangles. For each sub-population, the measures all nicely agree within the errors, thus indicating that their average value can be assumed as bona-fide density of the contaminating Galactic field population.

**Proper motions:** The two dataset discussed here (namely, the ACS/WFC-HST and GeMS/GSAOI data-sets) are separated by 6.3 years, a time-baseline sufficiently large to determine relative PMs. The approach we followed is the same already adopted in previous papers[42]. Here we describe only the main steps, while the detailed description will be presented in a forthcoming paper. First, we corrected the catalogues for geometric distortions by adopting literature prescriptions[43,44]. Then, for each star in each epoch we computed the mean (x,y) position in the images by averaging all the measures available in each data-set and by applying a σ-clipping rejection. As distortion-free reference frame, we adopted the catalogue obtained from the combination of the HST images. To derive the geometric transformations between the second and the first epoch, we selected a sample of likely cluster members along the RGB (master catalogue). We then applied a six-parameter linear transformation between the two epochs, as determined by using the stars in common between the GSAOI and the master catalogue. The derived transformations have been then applied to all the stars detected in each frame, by following the same procedure. The relative "first-pass" PMs were finally determined by measuring the difference of the mean (x,y) positions of the same star in the two epochs. We then repeated the entire procedure, by using as master catalogue the likely-member stars obtained from the first-pass PM derivation. Due to the dithering patterns adopted in the GeMS/GSAOI observations (aimed at covering the inter-chip gaps), the sample of stars for which PMs have been measured corresponds to 80% of the total.

**The extinction map in the direction of Liller1:** The construction of the extinction map in the direction of Liller 1 is described and discussed in detail in a dedicated paper (Cristina Pallanca et al., in preparation). Here we just briefly summarize the overall procedure. For each star in our catalogue we built a "local CMD" from a selected pool of N high-quality sources spatially lying in its vicinity (typically, we used a few dozens of sources within a distance of ~3"). To each star we then associated a differential extinction value determined as the shift along the reddening vector needed for the cluster mean ridge-line to match the "local CMD". Note that the extinction toward the inner Galaxy is expected[45] to be non-standard. Hence, appropriate absorption coefficients at different wavelengths have been adopted[46,47]. By assuming $A_\lambda = R_V \times c_\lambda \times E(B-V)$, here we adopted $R_V = 2.5$, instead of the canonical $R_V=3.1$, and $c_I=0.573$ and $c_K=0.107$. The values of extinction thus derived for each star have been used to construct the final reddening map in the direction of Liller 1, with an angular resolution better than 3". Our analysis confirms that the differential extinction in the cluster direction is large: $\delta A_I \sim 1$ mag and $\delta A_K \sim 0.2$ mag. The comparison between theoretical isochrones and the colour position of the RGB of the OP suggests a large value of the absolute extinction ($A_I=6.5$ mag and $A_K=1.2$ mag) that, together with the adopted absorption coefficients, correspond to an average colour excess $E(B-V) = 4.52\pm0.10$. Although this value is significantly larger than what found in the literature[16], the difference is essentially due to the lower value of $R_V$ adopted here. In fact, the extinction coefficients are fully comparable with previous determinations: for instance, by assuming[16] $E(B-V)=3.30$ and $R_V=3.1$, the resulting value of a $A_K$ (1.16) is fully comparable with ours.

**BSSs or not BSSs:** BSSs are anomalously massive (typically ~1.2 $M_\odot$)[48] core hydrogen-burning stars thought to form through the evolution of binaries and stellar collisions[49]. They have been found to be powerful tracers of the internal dynamical evolution of the parent stellar system[50-52]. Since they originate from sporadic and uncorrelated events, only limited numbers of BSSs are expected to be produced and, in fact, they usually appear as a sparse population in the CMD of star clusters: indeed only small numbers of BSSs are commonly observed (typically, less than 100-150 BSSs are found even in the most massive GCs[50], the richest population detected so far counting about 300 BSSs in ωCentauri[53]), and their fraction with respect to RGB and horizontal branch stars in the cluster central regions typically amounts to less than 10-15%. In Liller 1, the BP counts a population of stars comparable to that of the OP: we observe 2480 BP stars brighter than $I_{DRC}<24$, compared to 2916 stars at the same level of magnitude along the OP. Once corrected for completeness (see Section 'Artificial star experiments and completeness' below), these numbers becomes even more similar: 3030 and 3100, respectively. This strongly suggests that the BP is composed of a large population of young stars originated from a recent burst of star formation, instead of a population of objects (as BSSs) generated by sporadic events involving mass-transfer and collisions.

**The metallicity of Liller1.** The chemical composition of Liller1 is still poorly constrained and based on IR spectroscopy of a few bright giants only. High-resolution spectra with NIRSPEC at Keck[54] and with APOGEE[55] suggest values of [Fe/H] between ~1/4 solar and 0, with an average value of about half solar, and a mild (if any) α-enhancement. Low-resolution IR spectroscopy[56] of 8 bright giants also provide a similar value for the global metallicity. According to these measures we safely adopt [M/H]=−0.3.

**Artificial star experiments and completeness.** To investigate the completeness of the samples, we followed the standard procedure[57] of performing artificial star experiments. Briefly, artificial stars have been added to the acquired images by using the DAOPHOT II/ADDSTAR software. The artificial stars have been generated in the magnitude range 15<I<27 following the RGB mean ridge-line of the OP and the BP. In each run we simulated ~35,000 stars. To avoid the generation of "artificial crowding", the fake stars have been placed into the images following a regular grid of cells of approximately 15x15 pixels each, imposing that only one star per cell was simulated in each run. A total of more than 730,000 stars have been simulated in both in the ACS and the GeMS frames. The photometric analysis of the artificially-enriched frames has been executed following the same procedure used for the original frames (as described in Section `Data-set and Analysis' in Methods). The completeness of the observed sample for three different radial bins (0<r<15", 15"<r<30", and r>30") from the centre of gravity of Liller 1 is illustrated in Extended Data Figure 3, as function of both the I and K magnitudes. The large number of artificial star experiments also allowed us to estimate that the expected fraction of blends in the BP region is very small (~3 %).

**Measuring the age of the old stellar population.** In order to determine the age of the OP component, we first selected the portion of the CMD that is most sensitive to age variations: the SGB and the MS-turnoff region. Thus, stars tracing the OP have been selected in the $(I, I-K)_{DRC}$ CMDs at $23.5<I_{DRC}<25.3$, and isochrones[19,20] of different ages have been over-plotted adopting the quoted values of metallicity and distance modulus. For each isochrone we performed a $\chi^2$ analysis. The $\chi^2$ parameter has been computed by comparing the colour of each star to that of the isochrone at the same magnitude: $\chi^2 = \Sigma ((IK_{obs,j} - IK_{iso,j})^2/IK_{iso,j})$ where $IK_{obs,j}$ is the observed colour of star j, $IK_{iso,j}$ is the corresponding value of the colour read along the isochrones. By plotting the $\chi^2$ values obtained for isochrones of different ages (see Extended Data Figure 4), we found a well-defined minimum in correspondence of the 12 Gyr-old isochrones. The analysis indicates an uncertainty of ±1 Gyr. By taking into account also the uncertainties in the reddening and distance modulus, we finally adopted conservative estimate of 12±1.5 Gyr.

**Synthetic CMD.** Synthetic CMDs were constructed by using two sets of PARSEC isochrones[58]: one old model (t=12 Gyr) with metallicity Z=0.0048, and a sample of young isochrones with Z=0.03, in the age range 1-3 Gyr and separated by a regular step of 0.2 Gyr. Starting from these models, the synthetic populations were obtained by assuming a double power-law Initial Mass Function[59]. The colour and magnitude of the synthetic stars were randomly extracted from normal distributions with dispersion dependent on the photometric errors and differential-reddening residuals, as obtained from the observed catalogues. The star counts of the old synthetic population were then normalized to match the observed number of RGB stars in the magnitude range $22<I_{DRC}<23.5$. The age range and normalization factors adopted for the young synthetic populations were, instead, defined to minimize the difference in the colour and magnitude distributions between the synthetic and the observed populations in a regular grid of 0.5x0.5 mag cells in the ranges $21.4<I_{DRC}<24$ and $5.2<(I-K)_{DRC}<6.2$.

**Star excess in the RC region.** In order to estimate the expected number of RC stars belonging to the OP, we used the number of objects counted at the base of the RGB ($15<K_{DRC}<17$), where negligible contribution from the young component is expected because of the fast evolutionary time of this phase at such small ages. Theoretical tracks for stars of 0.9 $M_\odot$ (as appropriate for OP objects) indicates that the ratio between the evolutionary times in the RC and in this RGB magnitude range is approximately $t_{RC}/t_{RGB} = 0.9 \times 10^8 / 3.04 \times 10^8 = 0.3$. Hence, since the completeness-corrected counts (see Section 'Artificial star experiments and completeness' above) of RGB stars in the adopted magnitude range are 1175, we expect ~352 RC objects. By conservatively considering an additional 30%, due to the fact that RGB stars lying adjacent to the RC region appear indistinguishable from RC objects in the CMD, we expect a grand-total of ~457 stars belonging to the OP population in the RC region. We observe, instead, 765 stars. This clearly indicates that the young population is significantly contributing to the observed number counts, and suggests that the fraction of the young component, with respect to the total mass of Liller 1, can be as large as 40%, although detailed simulations are required to firmly assess this estimate.

**Data Availability:** The photometric data that support the plots and other findings of this study are available from the corresponding author upon reasonable request. The catalogs are also publicly downloadable from the web site of the Cosmic-Lab project (http://www.cosmic-lab.eu/Cosmic-Lab/Home.html). All the HST images are publicly available from the Mikulski Archive for Space Telescope (htttp://archive.stsci.edu/).

| Instrument | Date | Filters | Exposure time | Pixel size |
|---|---|---|---|---|
| ACS (HST) | August 17, 2019 | F606W (V) <br> F814W (I) | 6 x ~1300 s <br> 6 x ~840 s | 0.05"/pixel |
| GeMS (Gemini South) | April 20, 2013 | J <br> Kshort | 6 x 30 s <br> 6 x 30 s | 0.02"/pixel |
| WFC3 (HST) | March 1, 2019 | F606W (V) <br> F814W (I) <br> F110W (J) <br> F160W (H) | 2 x 1392 s, 2 x 1349 s <br> 6 x ~ 865 s <br> 1 x 1399 s <br> 1 x 1399 s | 0.04"/pixel <br><br> 0.14"/pixel |
| WFC3 (HST) | August 17, 2019 | F606W (V) <br> F814W (I) <br> F110W (J) <br> F160W (H) | 2 x 1392 s, 2 x 1349 s <br> 6 x ~ 865 s <br> 1 x 1399 s <br> 1 x 1399 s | 0.04"/pixel <br><br> 0.14"/pixel |
| NICMOS (HST) | June 27, 1998 | F110W (J) <br> F160W (H) | 1 x 224 s, 2 x 192 s <br> 1 x 160 s, 2 x 128 s | 0.2"/pixel |

**Extended Data Fig. 1** - Summary of the photometric dataset used in this work.

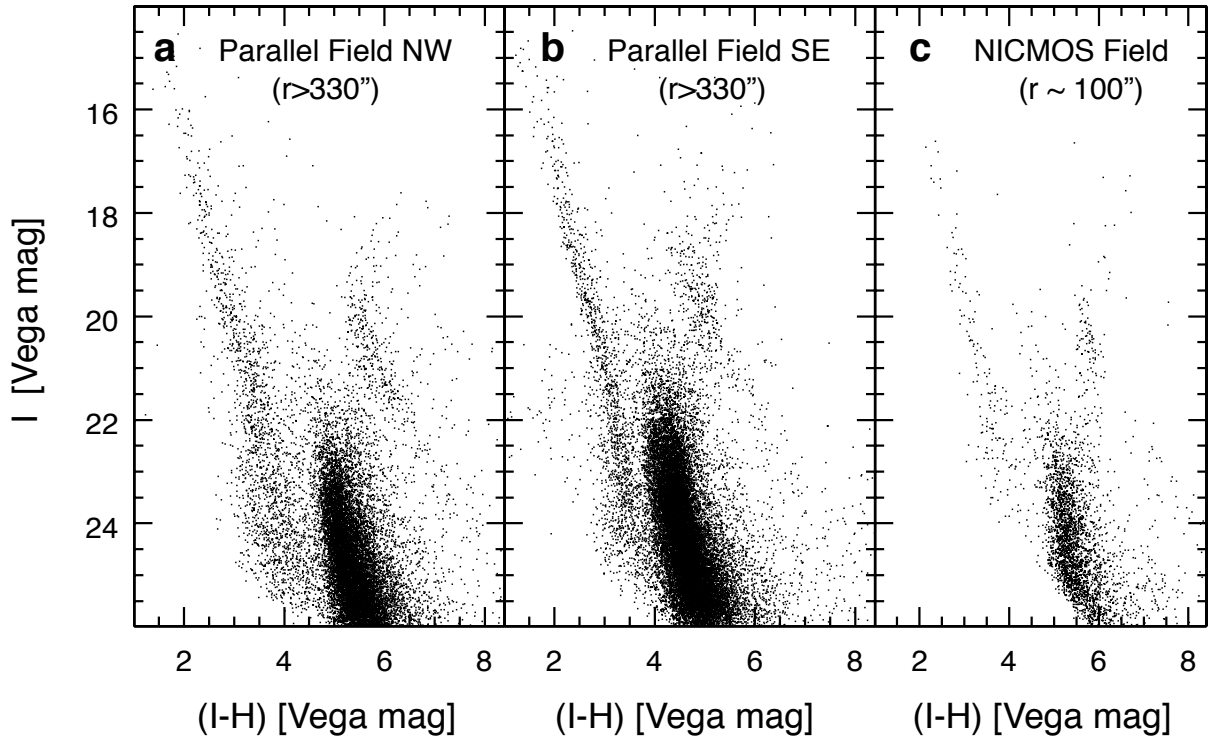

**Extended Data Fig. 2** - **a., b.** CMDs of the area sampled with the parallel WFC3 observations beyond the tidal radius of Liller1, illustrating the distribution of Galactic field stars in the vicinity of the cluster. **c.** CMD of a 50"x50" region approximately located at 100" from the centre of Liller1, obtained by combining the NICMOS observations retrieved from the Archive with the optical ACS-HST observations presented in this paper.

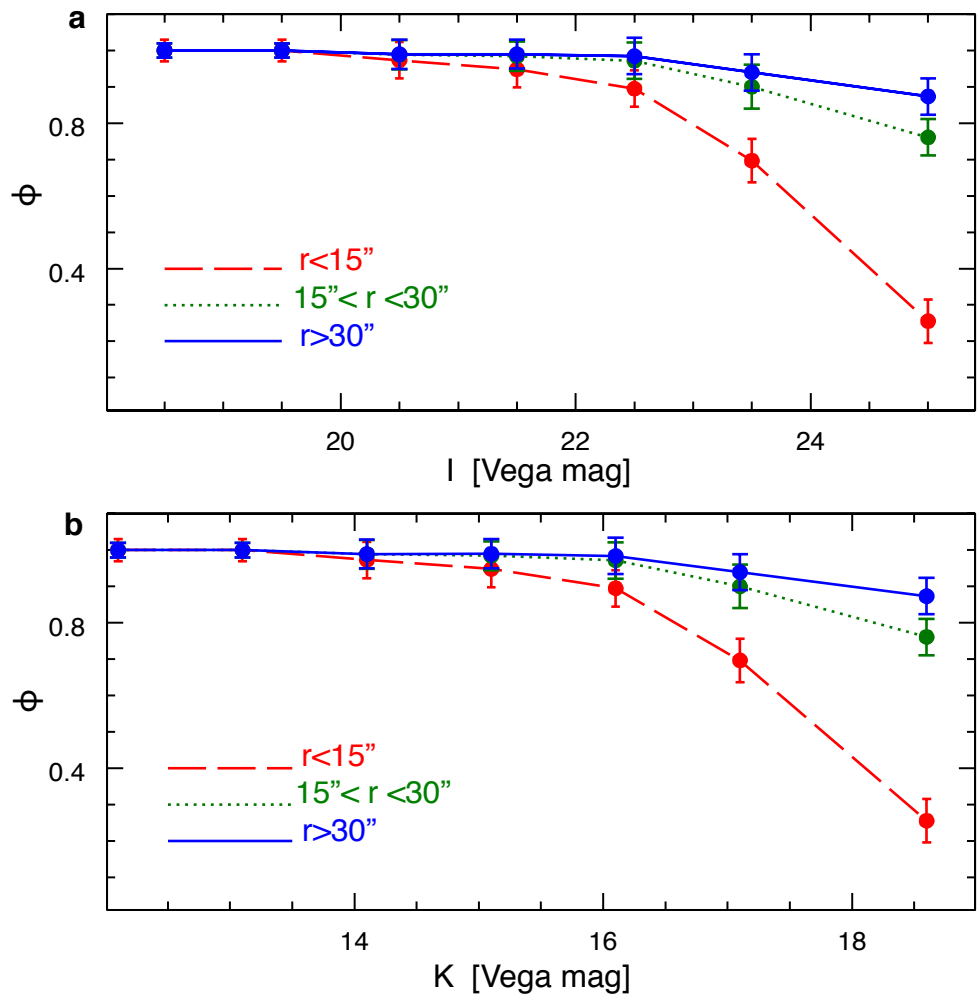

**Extended Data Fig. 3 - a., b.** The completeness of the sample is shown for three different radial bins from the gravitational centre of Liller 1 (see labels), both in the I-band and in the K-band (panels **a** and **b**, respectively).

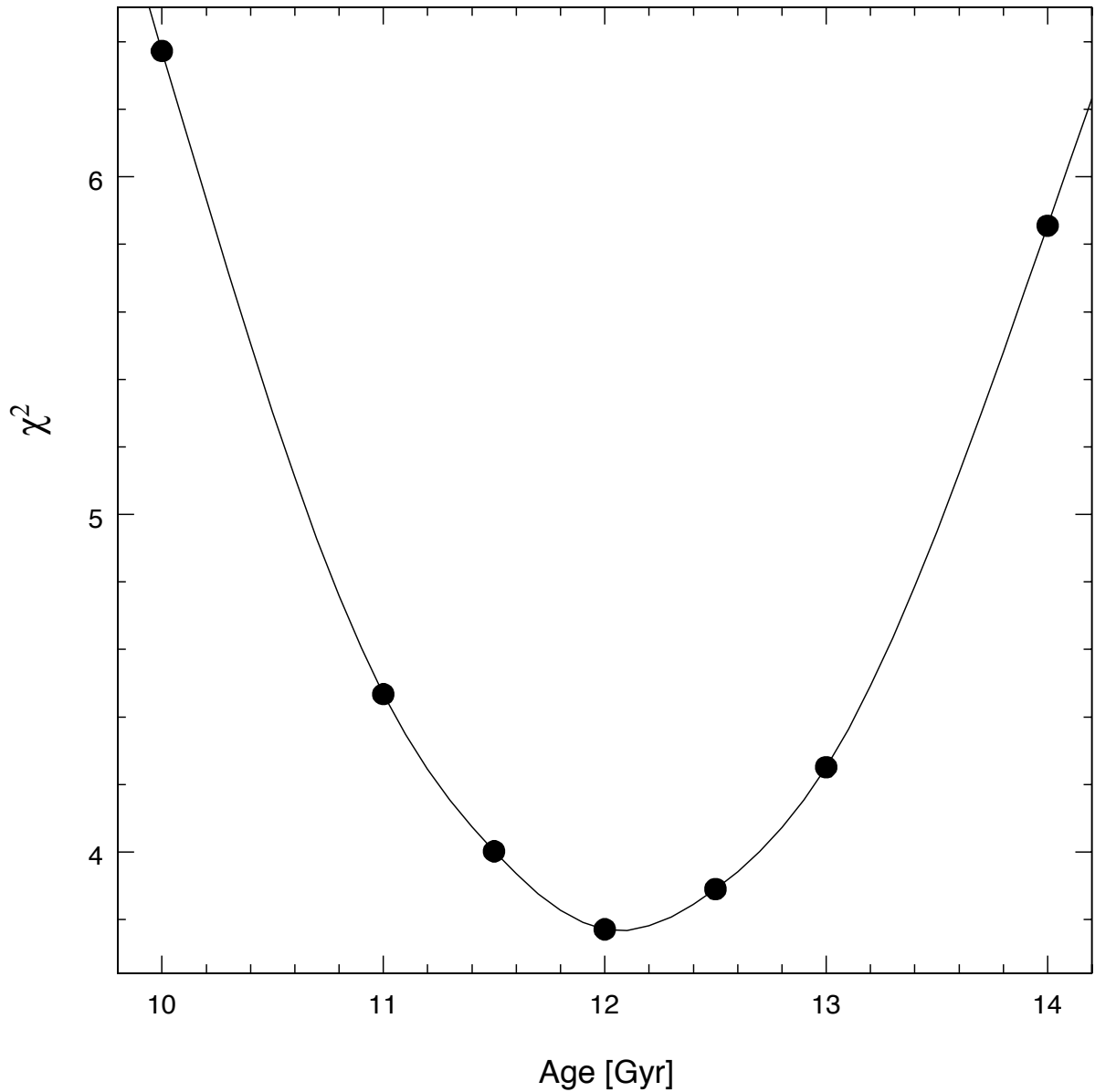

**Extended Data Fig. 4 - a., b.** The value of the χ2 parameter (see Section 'Measuring the age of the old stellar population' in Methods) is plotted as a function of the age of seven isochrones (with t=10, 11, 11.5, 12, 12.5, 13, 14 Gyr) computed for the quoted metallicity of the OP ([M/H]=−0.3). The minimum of the χ2 parameter suggests an age of 12 ± 1 Gyr for the OP of Liller 1.